\documentclass[aps,prb,twocolumn,showpacs,superscriptaddress,floatfix]{revtex4}
\usepackage{amssymb,amsmath}
\usepackage{graphicx}
\usepackage{color}

\begin{document}

\bibliographystyle{apsrev}

\title{Self-induced tunable transparency in layered superconductors}

\author{S.S.~Apostolov}
\affiliation{ A.Ya.~Usikov Institute for Radiophysics and
Electronics, National Academy of Sciences of Ukraine, 61085
Kharkov, Ukraine} \affiliation{V.N.~Karazin Kharkov National
University, 61077 Kharkov, Ukraine}

\author{Z.A.~Maizelis}
\affiliation{ A.Ya.~Usikov Institute for Radiophysics and
Electronics, National Academy of Sciences of Ukraine, 61085
Kharkov, Ukraine} \affiliation{V.N.~Karazin Kharkov National
University, 61077 Kharkov, Ukraine}

\author{M.A.~Sorokina}
\affiliation{V.N.~Karazin Kharkov National University, 61077
Kharkov, Ukraine}

\author{V.A.~Yampol'skii}
\affiliation{ A.Ya.~Usikov Institute for Radiophysics and
Electronics, National Academy of Sciences of Ukraine, 61085
Kharkov, Ukraine} \affiliation{V.N.~Karazin Kharkov National
University, 61077 Kharkov, Ukraine}\affiliation{Advanced Science
Institute, RIKEN, Saitama, 351-0198, Japan}

\author{Franco Nori}
\affiliation{Advanced Science Institute, RIKEN, Saitama, 351-0198,
Japan} \affiliation{Department of Physics, University of Michigan,
Ann Arbor, MI 48109, USA}

\begin{abstract}
We predict a novel nonlinear electromagnetic phenomenon in layered
superconducting slabs irradiated from one side by an
electromagnetic plane wave. We show that the reflectance and
transmittance of the slab can vary over a wide range, from nearly
zero to one, when changing the incident wave amplitude. Thus
changing the amplitude of the incident wave can induce either the total
transmission or reflection of the incident wave. In addition, the
dependence of the superconductor transmittance on the incident
wave amplitude has an unusual hysteretic behavior with jumps. This
remarkable nonlinear effect (self-induced transparency) can be
observed even at small amplitudes, when the wave frequency
$\omega$ is close to the Josephson plasma frequency $\omega_J$.
\end{abstract}





\maketitle

\section{Introduction.}

The recent growing interest in the unusual electrodynamic
properties of layered superconductors (see, e.g., the very recent review Ref.~\onlinecite{Thz-rev-2008-July}) is due to their possible applications,
including terahertz imaging, astronomy, spectroscopy, chemical and
biological identification. The experiments for the
$\mathbf{c}$-axis conductivity in high-$T_c$ $\rm
Bi_2Sr_2CaCu_2O_{8+\delta}$ single crystals justify the use of a
model in which the very thin superconducting CuO$_2$ layers are
coupled by the intrinsic Josephson effect through the thicker
dielectric
layers~\cite{Thz-rev-2008-July,Kl-Mu,Kl-Mu2,bran,plasma}. Thus, a
very specific type of plasma (the so-called, Josephson plasma) is
formed in layered superconductors. The current-carrying capability
of this plasma is strongly anisotropic, not only in the absolute
values of the current density. Even the physical nature of the
currents along and across layers is quite different: the current
along the layers is the same as in usual bulk superconductors,
whereas the current across the layers originates from the
Josephson effect. This Josephson current flowing along the
$\mathbf{c}$-axis couples with the electromagnetic field inside
the insulating dielectric layers, causing a specific kind of
elementary excitations called Josephson plasma waves (JPWs) (see,
e.g., Ref.~\onlinecite{Thz-rev-2008-July} and
references therein). Therefore, the layered structure of
superconductors favors the propagation of electromagnetic waves
through the layers.

The electrodynamics of layered superconductors is described by
\emph{nonlinear} coupled sine-Gordon
equations~\cite{SG-coupled,SG2,SG3,SG4,SG5,Thz-rev-2008-July}.
This nonlinearity originates from the nonlinear relation
$J\propto\sin\varphi$ between the Josephson interlayer current $J$
and the gauge-invariant interlayer phase difference $\varphi$ of
the order parameter. As a result of the nonlinearity, a number of
nontrivial nonlinear phenomena~\cite{nl1,nl2,nl3,nl4,nl5} (such as slowing
down of light, self-focusing effects, the pumping of weaker waves
by stronger ones, etc.) have been predicted for layered
superconductors. The nonlinearity plays a crucial role in the JPWs
propagation, even for small wave amplitudes, $|\varphi|\ll 1$,
when $\sin\varphi$ can be expanded into a series as
$(\varphi-\varphi^3/6)$, for frequencies close to the Josephson
plasma frequency.

In this paper, we predict a novel and unusual strongly nonlinear
phenomenon. The reflectance and transmittance of a superconducting
slab being exposed to an incident wave from one of its sides
depend not only on the wave frequency and the incident angle, but
also on the wave \emph{amplitude}. If the frequency $\omega$ of
irradiation is close to the Josephson plasma frequency $\omega_J$,
the transmittance of the slab can vary over a wide range,
\emph{from nearly zero to one}. Therefore, a slab of fixed
thickness can be absolutely transparent (when neglecting
dissipation) for waves of definite amplitudes, and nearly totally
reflecting for other amplitudes. This unusual property can be
described as \emph{self-induced tunable transparency}. Moreover,
the dependence of the transmittance on the wave amplitude shows
\emph{hysteretic} behavior with jumps. Tunable electromagnetically induced transparency is also now being studied in various contexts, including superconducting circuits and other lambda-type atoms (see, e.g., Refs.~\onlinecite{tt1,tt2}).

The large sensitivity of the slab transmittance to the wave
amplitude can be explained using a very clear physical
consideration. Let us consider a wave frequency $\omega$ which is
slightly smaller than the Josephson plasma frequency $\omega_J$.
In this case, \emph{linear} JPWs cannot propagate in the layered
superconductor (see, e.g., Ref.~\onlinecite{Thz-rev-2008-July}). So, the slab
is opaque for waves with small amplitudes, and the transmittance
is exponentially small due to the skin effect. According to
Refs.~\onlinecite{nl1,nl2,nl3}, the nonlinearity results in an effective
decrease of the Josephson plasma frequency and, thus,
\emph{nonlinear} JPWs with high-enough amplitudes can propagate in
the superconductor. Moreover, changing the amplitude of the
incident wave one can attain the conditions when the slab
thickness equals an integer number of half-wavelengths. Under such
conditions, the slab becomes transparent, with its transmittance
equal to one.

The paper is organized as follows. In section II, we discuss the
geometry of the problem and present the main equations for the
electromagnetic fields both in the vacuum and in the slab of
layered superconductor. In section III, we express the
transmittance $T$ in terms of the amplitude of the incident wave
and analyze this dependence in two cases: when the frequency
$\omega$ of the incident wave is either larger or smaller than the
Josephson plasma frequency $\omega_J$. In both cases, we study the
unusual hysteretic features of this dependence. The results of
numerical simulations support our theoretical predictions.

\section{Spatial distribution of the electromagnetic field}

\subsection{Geometry of the problem}

\begin{figure}
\begin{center}
\includegraphics*[width=8cm]{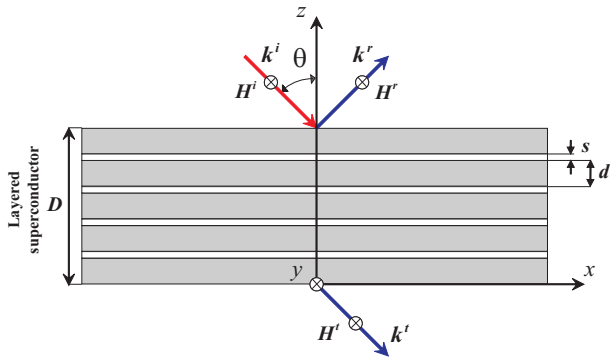}
\caption{(Color online) Geometry of the problem. A slab of layered
superconductor is irradiated from the upper side by a
$p$-polarized electromagnetic wave. }\label{SCSlab}
\end{center}
\end{figure}

We study a slab of layered superconductor of thickness $D$ (see
Fig.~\ref{SCSlab}). Superconducting layers of thickness $s$ are
interlayed by insulators of much larger thickness $d$ ($s\ll d$).
We assume the number of layers to be large, allowing the use of
the continuum limit, and not to consider the spatial distribution
of the electromagnetic field inside each layer. The coordinate
system is chosen in such a way that the crystallographic
$\mathbf{ab}$-plane coincides with the $xy$-plane, and the
$\mathbf{c}$-axis is along the $z$-axis. The plane $z=0$
corresponds to the lower surface of the slab.

A monochromatic electromagnetic plane wave of frequency $\omega$
is incident on the upper surface of the slab, and it is partly
reflected and partly transmitted through the slab. We consider the
incident wave of the transverse magnetic polarization, when the
magnetic field is parallel to the surface of the slab,
\begin{equation}\label{e1}
{\vec E} = \{E_x, 0, E_z\}, \qquad {\vec H} = \{0, H, 0\}.
\end{equation}
The incident angle $\theta$ is considered to be not close to zero,
so that both components $k_x$ and $k_z$ of the wave-vector
$\textbf{k}_i$ are of the order of $\omega/c$.

\subsection{Electromagnetic field in the vacuum}

The magnetic field $H^{u}$ in the upper vacuum semispace ($z>D$)
can be represented as a sum of the incident and reflected waves
with amplitudes $H_0$ and $H_r$, respectively. The field $H^{l}$
in the vacuum semispace below the sample ($z<0$) is the
transmitted wave with amplitude $H_t$. The upper ($H^{u}$) and
lower ($H^{l}$) fields and  can be written in the following form,
\[
H^{u}=H_0\cos[k_xx-\omega t-k_z (z-D)]
\]
\begin{equation}\label{e2}
+H_r\cos [k_xx-\omega t+k_z (z-D)+\alpha],
\end{equation}
\begin{equation}\label{h vac down}
H^{l}=H_t \cos (k_xx-\omega t-k_z z+\beta).
\end{equation}
Here
\begin{equation}\label{e3}
k_x=\frac{\omega}{c}\sin\theta,\quad k_{z}=\frac{\omega}{c}
\cos\theta,
\end{equation}
are the components of the wave-vector $\textbf{k}_i$ of the
incident wave, $\alpha$ and $\beta$ are the phase shifts of the
reflected and transmitted waves. Using Maxwell equations one can
derive the electric field components in the vacuum:
\[
E_x^{u}=-H_0\cos\theta\cos[k_xx-\omega t-k_z (z-D)]
\]
\begin{equation}\label{el1x}
+H_r\cos\theta\cos [k_xx-\omega t+k_z (z-D)+\alpha],
\end{equation}
\[
E_z^{u}=-H_0\sin\theta\cos[k_xx-\omega t-k_z (z-D)]
\]
\begin{equation}\label{el1z}
-H_r\sin\theta\cos [k_xx-\omega t+k_z (z-D)+\alpha],
\end{equation}
\begin{equation}\label{el2x}
E_x^{l}=-H_t\cos\theta \cos (k_xx-\omega t-k_z z+\beta),
\end{equation}
\begin{equation}\label{el2}
E_z^{l}=-H_t\sin\theta \cos (k_xx-\omega t-k_z z+\beta).
\end{equation}

\subsection{Electromagnetic field in the layered superconductor}

The electromagnetic field inside a layered superconductor slab is
determined by the distribution of the gauge-invariant phase
difference $\varphi(x,z,t)$ of the order parameter between the
layers (see, e.g., Ref.~\onlinecite{Thz-rev-2008-July}),
\[
\frac{\partial H^s}{\partial x}=\frac{{\cal
H}_0}{\lambda_c}\Big(\frac{1}{\omega_J^2}\frac{\partial^2\varphi}{\partial
t^2}+\sin\varphi\Big),
\]
\begin{gather}\label{sfield}
E^s_x=-\frac{\lambda^2_{ab}}{c}\frac{\partial^2 H^s}{\partial
z\partial t},  \qquad E_z^s=\frac{{\cal H}_0
\lambda_c}{c}\frac{\partial\varphi}{\partial t}.
\end{gather}
Here ${\cal H}_0=\Phi_0/2\pi d\lambda_c$, $\Phi_0=\pi c\hbar/e$ is
the magnetic flux quantum, $\lambda_{ab}$ and
$\lambda_c=c/\omega_J\epsilon^{1/2}$ are the London penetration
depths across and along the layers, respectively. The Josephson
plasma frequency is defined as
\begin{equation}\label{e7}
\omega_J = \sqrt{\frac{8\pi e d J_c}{\hbar\epsilon}},
\end{equation}
where $J_c$ is the critical value of the Josephson current
density, $\epsilon$ is the permittivity of the dielectric layers
in the slab. We omit the relaxation terms because, at low
temperatures, they do not play an essential role in the phenomena
considered here.

The phase difference $\varphi$ obeys a set of coupled sine-Gordon
equations, that, in the continuous limit, takes on the following
form (see, e.g., Ref.~\onlinecite{Thz-rev-2008-July} and
references therein):
\begin{equation}\label{SGbasic}
\left(1-\lambda_{ab}^2\frac{\partial^2}{\partial
z^2}\right)\left[\frac{1}{\omega^2_J}\frac{\partial^2
\varphi}{\partial t^2}  +\sin\varphi \right] -
\lambda_c^2\frac{\partial^2 \varphi}{\partial x^2}=0.
\end{equation}

In this paper, we study the case of weak nonlinearity: when the
Josephson current density $J_c \sin \varphi$ can be expanded as a
series for small $\varphi$, up to the third order, $J_c \sin
\varphi \approx J_c(\varphi-\varphi^3/6)$. We consider frequencies
$\omega$ close to $\omega_J$ and introduce a dimensionless
frequency,
\[
\Omega=\frac{\omega}{\omega_J},
\]
close to one. In this case, in spite of the weakness of the
nonlinearity in Eq.~(\ref{SGbasic}), the \emph{linear terms nearly
cancel each other}, and the term $\varphi^3$ plays a crucial role
in this problem. Moreover, when the frequency $\omega$ is close to
the Josephson plasma frequency $\omega_J$, one can neglect the
generation of higher harmonics~\cite{nl1,nl3}.

It should be also noted that the nonlinearity provides an
effective decrease of $\omega_J$. Indeed, the expression in the
square brackets in Eq.~(\ref{SGbasic}) can be presented in the form
$[(\omega^{\rm eff}_J)^{-2}\partial^2 \varphi/\partial t^2  +\varphi]$,
where
\[
\omega^{\rm eff}_J \approx \omega_J\Big(1-\dfrac{\varphi^2}{12}\Big).
\]
For not very small $\varphi$, the frequency of
the incident wave can be greater than the effective Josephson
plasma frequency $\omega^{\rm eff}_J$ and, therefore, the nonlinear Josephson plasma waves can propagate
across the superconducting layers.

The $z$-component of the electric field induces a charge in the
superconducting layers when the charge compressibility is finite.
This results in an additional interlayer coupling (so-called,
capacitive coupling). Such a coupling significantly affects the
properties of the \emph{longitudinal} Josephson plasma waves with
wave-vectors perpendicular to the layers. The dispersion equation
for linear Josephson plasma waves with arbitrary direction of the
wave-vectors, taking into account capacitive coupling, was derived
in Ref.~\onlinecite{helm3}. According to this dispersion equation, the
capacitive coupling can be safely neglected in our case, when the
wave-vector has a component $k_x \sim \omega/c$ along the layers,
due to the smallness of the parameter $\alpha = R_D^2\epsilon/sd$.
Here $R_D$ is the Debye length for a charge in a superconductor.

We seek a solution of Eq.~(\ref{SGbasic}) in the form of a wave
running along the $x$-axis,
\begin{equation}\label{e8}
\varphi (x,z,t)= a(z)|1-\Omega^2|^{1/2}\sin\big[k_xx - \omega
t+\eta(z)\big],
\end{equation}
with the $z$-dependent amplitude $a$ and phase $\eta$. We
introduce the dimensionless $z$-coordinate,
\begin{equation}\label{zeta-delta}
\zeta=\frac{\kappa z}{\lambda_{ab}},\quad \kappa=\frac{\lambda_c
k_x}{|1-\Omega^2|^{1/2}},
\end{equation}
and the normalized thickness of the sample
$\delta=D\kappa/\lambda_{ab}$.

Substituting the phase difference $\varphi$ given by
Eq.~\eqref{e8} into Eq.~\eqref{SGbasic}, one obtains two
differential equations for the functions $\eta(\zeta)$ and
$a(\zeta)$. The first of them is
\begin{equation}\label{eta}
\eta'(\zeta)=\frac{L}{h^2(\zeta)}
\end{equation}
where $L$ is an integration constant, prime denotes derivation
over $\zeta$, and
\begin{equation}\label{h(a)}
h(\zeta)=\pm a(\zeta) - \frac{a^3(\zeta)}{8}.
\end{equation}
The sign in this equation is plus for $\Omega<1$ and minus for
$\Omega>1$, i.e., it is opposite to the sign of the following
important parameter, the
\[
{\rm frequency \,\, detuning} = \Omega-1.
\]

The coupled sine-Gordon equations \eqref{SGbasic} give also the
differential relation for $h(\zeta)$:
\begin{equation}\label{sine-Gordon}
h''=a+\frac{L^2}{h^3}-\frac{h}{\kappa}.
\end{equation}
Equations~\eqref{e8}, \eqref{eta}, \eqref{h(a)},
and~\eqref{sine-Gordon} allow one to calculate the distribution of
the phase difference $\varphi (x,z,t)$ and then, using
Eq.~(\ref{sfield}), the electromagnetic field inside the
superconducting slab.

\section{Transmittance of the superconducting slab}

\subsection{Main equations}

In this section, we analyze the transmittance $T$ of a slab of
layered superconductor. We rewrite the expressions for the
magnetic field $H^s$ and for the $x$-component $E_x^s$ of the
electric field inside the slab using Eqs.~\eqref{sfield} and
Eq.~\eqref{e8},
\[
H ^s(x,\zeta,t)=-{\cal H}_0
\frac{|1-\Omega^2|}{\kappa}h(\zeta)\cos(k_xx - \omega
t+\eta(\zeta)\big),
\]
\[
E_x ^s(x,\zeta,t)={\cal H}_0
\Gamma\frac{|1-\Omega^2|\cos\theta}{\kappa } \,
\]
\begin{equation}\label{sfield_h}
\times \left[h(\zeta)\sin\big(k_xx - \omega t+\eta(\zeta)\big)
\right]'.
\end{equation}
Here we introduce the parameter
\[
\Gamma=\frac{\lambda_{ab}\kappa}{\lambda_c\sqrt{\epsilon}\cos\theta},
\]
which is usually small for layered superconductors. Now we can
find the unknown amplitudes of the reflected and transmitted waves
by matching the magnetic fields and the $x$-components of the
electric field at both interfaces (at $z=0$ and $z=D$) between the
vacuum and the layered-superconductor. Using Eqs.~\eqref{sfield_h}
for the fields in the superconductor and Eqs.~\eqref{e2}, \eqref{h
vac down}, \eqref{el1x}, and \eqref{el2x} for the fields in the
vacuum, we obtain the following three equations for the amplitudes
$a(0)$, $a(\delta)$ and their derivatives on both surfaces of the
layered superconductor:
\begin{gather}
\bigg(h(\delta)+\frac{\Gamma
L}{h(\delta)}\bigg)^2+\Gamma^2\big[h'(\delta)\big]^2
=4h_0^2,\label{upper}\\
h^2(0)=\Gamma L,\label{lower ampl}\\
a'(0)=0.\label{lower deriv}
\end{gather}
Here
\begin{equation}\label{def_h_0}
h_0=\frac{H_0}{{\cal H}_0} \frac{\kappa}{|1-\Omega^2|}
\end{equation}
is the normalized amplitude of the incident wave. These three
equations, together with the coupled sine-Gordon
equations~\eqref{sine-Gordon}, determine the integration constant
$L$ for each amplitude of the incident wave $h_0$ (see
Eq.~\eqref{eta}). It is important to note that the constant $L$
defines directly the transmittance $T$ of the superconducting
slab. Indeed, according to Eq.~\eqref{lower ampl}, we have
\begin{equation}\label{R}
T=\frac{h^2(0)}{h_0^2}=\frac{\Gamma }{h_0^2}L.
\end{equation}

The nonlinearity of Eq.~\eqref{sine-Gordon} leads to the
\emph{multivalued} dependence of the transmittance on the
amplitude of the incident wave. In the following subsections, we
analyze this unusual dependence, for both cases of negative
($\Omega<1$) and positive ($\Omega>1$) frequency detunings.

\subsection{Transmittance of a superconducting slab for $\omega<\omega_J$}


We start with the case when the frequency of the incident wave is
smaller than the Josephson plasma frequency. In this frequency
range, the linear Josephson plasma waves cannot propagate in
layered superconductors. This corresponds to an exponentially
small transmittance in the slab, due to the skin effect. However,
the nonlinearity promotes the wave propagation because of the
effective decrease of the Josephson plasma frequency.

\begin{figure}
\begin{center}
\includegraphics*[width=8cm]{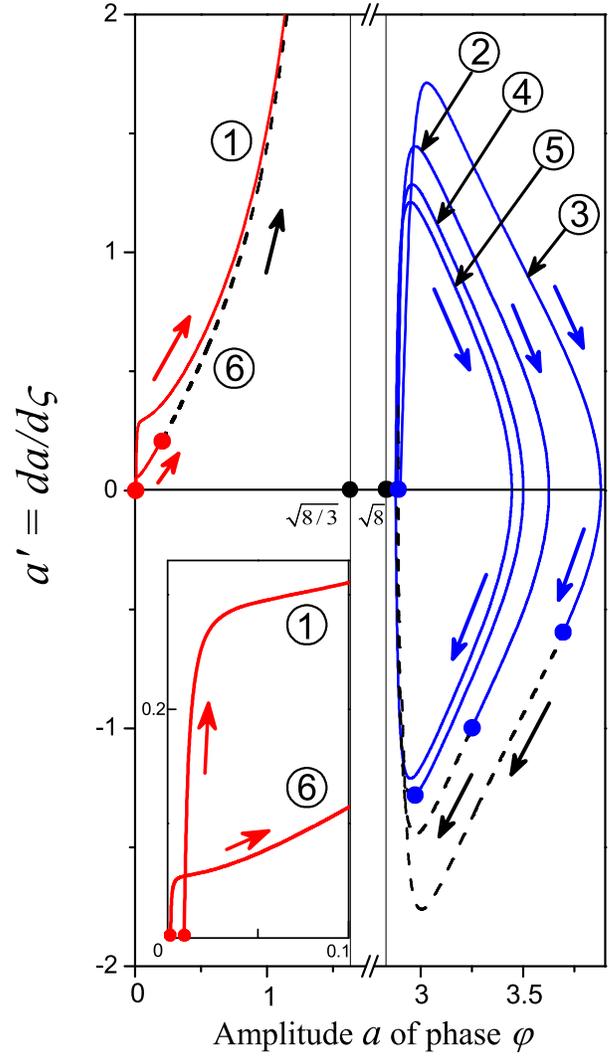}
\caption{(Color online) Phase trajectories $a'(a)$ for the
\emph{negative} frequency detuning $\Omega-1= - 5 \cdot 10^{-5}$.
The numbers near the curves correspond to the same numbers of
points in the $T(h_0)$ plot shown by the solid curve in
Fig.~\ref{fig t(h) omega<1}. The movement along the phase
trajectories in the directions shown by the arrows corresponds to
the growth of the spatial coordinate $\zeta$ (proportional to
$z$), from zero to $\delta$, inside the slab. The solid lines show
the portions of the phase trajectories that correspond to $0<\zeta
<\delta$. The lower and upper surfaces of the slab correspond to
the solid circles on the trajectories. The inset in the bottom
left shows the enlarged region near the point $(a=0,\, a'=0)$. The
other parameters are the same as for Fig.~\ref{fig t(h) omega<1}.}
\label{phase diagr omega<1}
\end{center}
\end{figure}

Solving Eq.~\eqref{sine-Gordon} with the boundary
conditions~\eqref{upper},~\eqref{lower ampl}, and~\eqref{lower
deriv} one can find the constant $L$ and then calculate the
transmittance using Eq.~\eqref{R}. Figure~\ref{fig t(h) omega<1}
demonstrates the numerically-calculated dependence of the
transmittance $T$ on the normalized amplitude $h_0$ of the
incident wave. To analyze this dependence, we consider the spatial
distribution of the gauge-invariant phase difference $\varphi$ of
the order parameter and the phase trajectories $a'(a)$. We show
these curves $a'(a)$ in Fig.~\ref{phase diagr omega<1}. An
increase of the spatial coordinate $\zeta$ [which is essentially
$z$, as defined in Eq.~\eqref{zeta-delta}] from zero to $\delta$
corresponds to moving along the phase trajectory $a'(a)$. The
point $\zeta=0$ (i.e., $z=0$) corresponds to the starting point on
the phase trajectory $a'(a)$. According to Eq.~\eqref{lower
deriv}, all phase trajectories start from the points where $a'=0$.
Different trajectories in $a'$ versus $a$ can be characterized by
the values of $a(0)$ in these starting points. Each trajectory
corresponds to some value of the normalized amplitude $h_0$ of the
incident wave, and, according to Eqs.~\eqref{h(a)}, \eqref{lower
ampl}, and \eqref{R}, the value of $a(0)$ \emph{defines} the
constant $L$ and the transmittance of the slab.

\begin{figure}
\begin{center}
\includegraphics*[width=8cm]{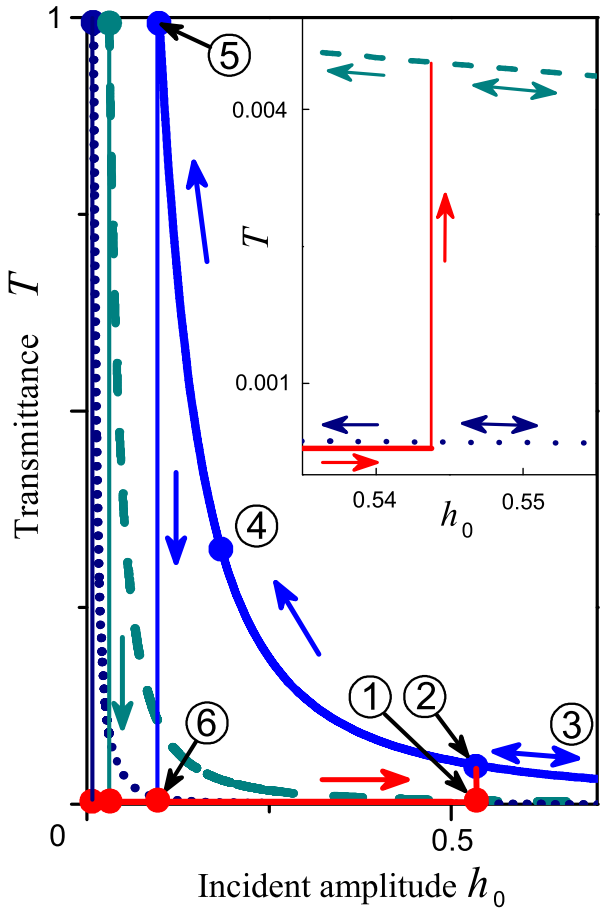}
\caption{(Color online) Dependence of the transmittance $T$ [see
Eq.~\eqref{R}] on the normalized amplitude $h_0$ of the incident
magnetic field [see Eq.~\eqref{def_h_0}] for different values of
the \emph{negative} frequency detuning: $(\Omega-1)= - 5 \cdot
10^{-5}$ (solid curve), $- 5 \cdot 10^{-4}$ (dashed curve), $- 5
\cdot 10^{-3}$ (dotted curve). Arrows show the change of the
transmittance when changing $h_0$. The numbers near the points on
the solid $T(h_0)$ curve correspond to the same numbers of the
phase trajectories $a'(a)$ shown in Fig.~\ref{phase diagr
omega<1}. The inset shows the enlarged region near the point 1.
The values of the parameters are: $\delta=2$, $\lambda_c=4\cdot
10^{-3}$~cm, $\lambda_{ab}=2000$~\AA,
$\omega_J/2\pi=0.3$~THz, and $\theta=45^\circ$.} \label{fig t(h)
omega<1}
\end{center}
\end{figure}

The low-amplitude (quasi-linear) branch of the $T(h_0)$ dependence
ranges within the interval $0<h_0<(8/27)^{1/2}$ of the amplitudes
of the incident waves. This branch is shown in Fig.~\ref{fig t(h)
omega<1} by the red solid curve close to the abscissa. For small
amplitudes $h_0\ll 1$, we deal with a linear problem, when the
phase difference $\varphi$ and the electromagnetic field in the
superconductor can be found in the form of linear combinations of
exponential functions of $z$. In this case, the transmittance $T$
can be found asymptotically for small $\Gamma$,
\begin{equation}\label{lin}
T(h_0\ll1)\approx
\frac{4\Gamma^4}{\sinh^2\left[\delta(1-\kappa^{-2})\right] +
4\Gamma^4}, \quad \Gamma \ll 1.
\end{equation}
This transmittance is very close to zero regardless of the
frequency detuning $(\Omega - 1)$. As we will see below, the
``sinh'' above, for $\omega < \omega_J$, will become ``sin'' for
$\omega > \omega_J$.

The phase trajectories that correspond to the low-amplitude
solutions occupy the region $a<(8/3)^{1/2}$. For small $h_0$,
these trajectories are close to the point $(a=0,\,a'=0)$ (as an
example of such trajectory, see curve 6 in Fig.~\ref{phase diagr
omega<1}). An increase of the amplitude $h_0$ leads to the growth
of the length of the phase trajectory and this length tends to
infinity when $h_0 \rightarrow (8/27)^{1/2}$ (see curve 1 in
Fig.~\ref{phase diagr omega<1}).

The high-amplitude branches of the $T(h_0)$ dependence correspond
to the solutions with $a(\zeta)>{8}^{1/2}$. Such branches are
shown in Fig.~\ref{fig t(h) omega<1} by dotted, dashed, and blue
solid curves for different values of the frequency detuning. The
high-amplitude solutions describe nonlinear Josephson plasma waves
that can propagate in the layered superconductor even for negative
frequency detuning (for $\Omega < 1$). The corresponding phase
trajectories are closed curves (see the closed curves in
Fig.~\ref{phase diagr omega<1} for $a> {8}^{1/2}$). Note that the value of $h$ is negative for
$a>8^{1/2}$. For this case, we can consider $h$ to be positive, but the phase of the
incident wave must be shifted by $\pi$.

The oscillating character of the high-amplitude solutions results
in much higher values of the transmittance, compared to the case
of exponentially-small quasi-linear solutions. As seen in
Fig.~\ref{fig t(h) omega<1}, the transmittance varies over a wide
range, from nearly zero to one, depending on the amplitude $h_0$
of the incident magnetic field. It is important to note that the
wavelengths of the nonlinear waves in the superconductor depend
strongly on the incident wave amplitude $h_0$. So, changing $h_0$
one can control the relation between the wavelength and the
thickness of the slab. The transmittance is very sensitive to this
relation, and one can attain total transparency of the slab
choosing the optimal value $h_0^{\rm max}$ of the amplitude $h_0$.

For high enough amplitudes $h_0$, the sample thickness $D$ is
larger than the half-wavelength. In this case, the change of the
coordinate $\zeta$ in the interval $0<\zeta<\delta$ corresponds to
the movement along a section of the phase trajectory loop (see the
trajectories 2, 3, 4 in Fig.~\ref{phase diagr omega<1}). When
decreasing $h_0$, the wavelength increases, the movement along the
phase trajectory approaches the complete loop, and the
transmittance of the slab increases. Finally, for a specific value
of $h_0$, the wavelength becomes equal to the sample thickness,
the phase trajectory draws a complete loop, and the transmittance
becomes equal to one (see the trajectory 5 in Fig.~\ref{phase
diagr omega<1} and point 1 in Fig.~\ref{fig t(h) omega<1}).

\begin{figure}
\begin{center}
\includegraphics*[width=8cm]{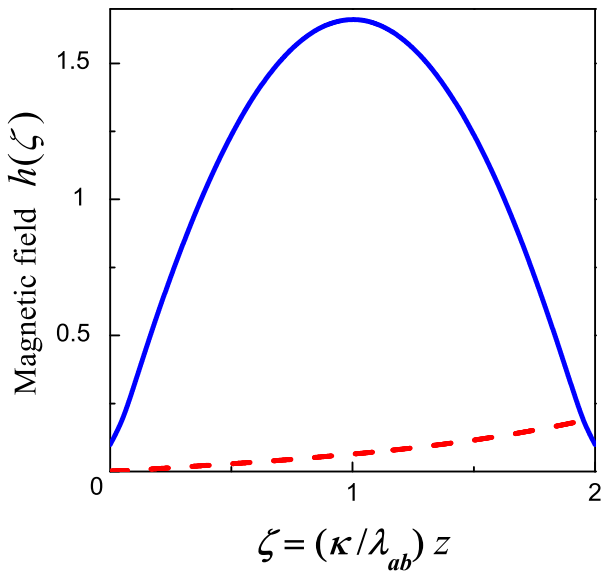}
\caption{(Color online) Spatial distribution of the amplitude $h$
of the magnetic field \emph{inside} the superconducting plate.
Solid and dashed curves are plotted for the points 5 and 1 shown
in Fig.~\ref{fig t(h) omega<1}, respectively. Point 1 corresponds
to the low-amplitude branch of the $T(h_0)$ dependence, when the
transmission coefficient is close to zero, and the amplitude $h$
of the magnetic field near the lower interface ($z=0$) of the slab
is exponentially small. The transmittance $T$ for the point 5 is
equal to one, and the amplitudes of the fields near the upper and
lower interfaces of the slab coincide. Here, the frequency
detuning is $(\Omega -1)= -5 \cdot 10^{-5}$, and the other
parameters are the same as for Fig.~\ref{fig t(h) omega<1}.}
\label{fig_distribution}
\end{center}
\end{figure}

The amplitude dependence of the transmittance can be found
asymptotically for small values of the parameter $\Gamma$ and for
not very thick slabs, $\delta\lesssim 1$,
\begin{equation}\label{h0igla}
T(h_0) \, \simeq \,
\frac{4\Gamma^2}{\delta^2}\Big(\frac{\delta^2}{\sqrt{2}h_0}+1\Big)^2.
\end{equation}
In the case of total transparency ($T=1$) of the slab, for
$h_0^{\rm max}\simeq 2^{1/2}\Gamma\delta$, both the electric and
magnetic fields take on the same values on the upper and lower
surfaces of the slab. Thus, the amplitudes of the incident and
transmitted waves are equal. The corresponding spatial
distribution of the magnetic field is shown by the solid curve in
Fig.~\ref{fig_distribution}.

The nontrivial feature of the $T(h_0)$ dependence can be seen from
its hysteretic behavior with jumps. Let the amplitude $h_0$ of the
incident wave increase from zero. In this case, the transmittance
is close to zero, and the $T(h_0)$ dependence follows the
low-amplitude branch shown by the red solid line near the abscissa
in Fig.~\ref{fig t(h) omega<1}. When the amplitude reaches the
critical value $(8/27)^{1/2}$ (point 1), further movement along
this branch is impossible, and a jump to point 2, on the
high-amplitude branch, occurs. A further increase in the amplitude
$h_0$ results in a monotonic decrease of the transmittance.

Afterwards, if the amplitude $h_0$ starts to decrease, then the
$T(h_0)$ dependence does not follow the same track. Indeed, when
the point 2 is passed, the transmittance continues to follow the
high-amplitude branch. Decreasing the amplitude $h_0$ results in a
further increase of the transmittance. When it becomes equal to
one (point 5), it is not possible to continue the movement along
the high-amplitude branch, and a return jump to the low-amplitude
branch occurs.

It should be noted that the jump from the low-amplitude
branch (which corresponds to the exponentially small transmittance) to the high-amplitude one
(with much higher transmittance) can be observed when changing the wave frequency
$\omega$ for the constant amplitude $H_0$ of the incident wave.
This jump occurs when the frequency detuning $(1-\Omega)$ becomes equal to the threshold value
\[
\left(1-\Omega_{\rm cr}\right)=\dfrac{3}{4}\left(\lambda_c k_x \dfrac{H_0}{\mathcal{H}_0}
\right)^{2/3}.
\]

\subsection{Mechanical analogy}

The problem discussed in this paper has a deep and very interesting mechanical analogy. Indeed,
Eqs.~\eqref{sine-Gordon} and Eq.~\eqref{eta} describe a motion of a particle with unite mass in a centrally symmetric potential. The amplitude $h(\zeta)$ of the magnetic field, the phase $\eta(\zeta)$, and the coordinate $\zeta$ across the layers of the superconductor play the roles of the radial coordinate of the particle, its polar angle, and time, respectively. Moreover, the constant $L$ in Eqs.~\eqref{sine-Gordon} and Eq.~\eqref{eta} can be regarded as the conserved angular momentum of the particle.

Integrating Eq.~\eqref{sine-Gordon} for the radial motion of the particle, we obtain the energy conservation law for the particle,
\begin{equation}\label{energy}
\dfrac{(h')^2}{2}+ U_{\rm eff}(h)={\cal E},
\end{equation}
with the effective potential
\begin{equation}\label{ueff}
U_{\rm eff} (h)=\dfrac{L^2}{2h^2}+\dfrac{h^2}{2\kappa}-\int^h a(\tilde{h}) \,
{\rm d}\tilde{h}.
\end{equation}
The first term in Eq.~\eqref{energy} describes the kinetic energy of the radial motion of the particle, ${\cal E}$ is the total energy of the particle. The first term in Eq.~\eqref{ueff} is the centrifugal energy and the last two terms represent the potential of the central field.

\begin{figure}
\begin{center}
\includegraphics*[width=8cm]{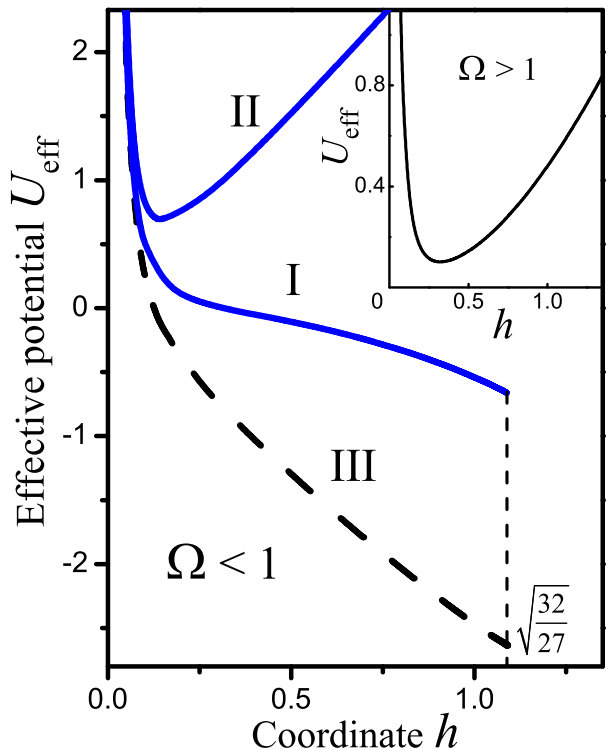}
\caption{(Color online) Dependence of the effective potential
$U_{\rm eff}$ defined by Eq.~\eqref{ueff} on the radial coordinate
$h$. The movement of the particle in this potential represents the mechanical analog for the spatial
distribution of the amplitude $h$ of the magnetic field in the superconductor. The main panel
shows curves I, II, and III that correspond to the three branches of the $a(h)$ dependence
for the case of negative frequency detuning, $\Omega<1$. The inset shows the $U_{\rm eff}(h)$ curve for the opposite case, $\Omega>1$, when only one branch of the $a(h)$ dependence exists. The value of constant $L$ is $0.1$.} \label{fig_ueff_<1}
\end{center}
\end{figure}

The plot of the dependence $U_{\rm eff} (h)$ is shown in Fig.~\ref{fig_ueff_<1} for the case of negative detuning ($\Omega < 1$). This dependence is three-valued and corresponds to the three branches of the function $a(h)$ (see Eq.~\eqref{h(a)}). Thus, the multivaluedness of the dependence $a(h)$ results in the multivaluedness of the effective potential $U_{\rm eff} (h)$ and, therefore, there exist several possibilities for the particle motion. In terms of our electrodynamical problem, this means that several field distributions in the superconductor can be realized for the same amplitude $h_0$ of the incident wave.

Curve I in Fig.~\ref{fig_ueff_<1} shows the potential that corresponds to the low-amplitude solutions of our electrodynamical problem. The motion of the particle (from right to left) in this potential is monotonic that corresponds to the monotonic decrease of the field deep into the superconductor. According to Eq.~\eqref{lower deriv}, the stop-point of the particle ($h'=0$) corresponds to the lower boundary of the superconductor.

Since curve I is terminated in the point $h=(32/27)^{1/2}$, it cannot define the particle motion for high enough $h$. In this case, the particle moves in the potential described by curve II in Fig.~\ref{fig_ueff_<1}. This motion is finite and periodical. It corresponds to the high-amplitude solutions of our electrodynamical problem. Curve III in Fig.~\ref{fig_ueff_<1} represents a branch of the $U_{\rm eff}(h)$ dependence which cannot be realized when changing the amplitude $h_0$ of the incident wave.

\subsection{Transmittance of a superconducting slab for $\omega>\omega_J$}

Now we study the transmittance of a slab of layered
superconductors for waves with frequencies higher than the
Josephson-plasma frequency, $\Omega>1$. Contrary to the case
$\Omega<1$, even linear Josephson plasma waves can propagate in
the layered superconductor. Therefore, the transmittance is not
exponentially small and can vary over a wide range, depending on
the relation between the wavelength and the thickness of the slab:
\begin{equation}\label{linear}
T(h_0\ll1)\approx
\frac{4\Gamma^4}{\sin^2\left[\delta(1-\kappa^{-2})\right] +
4\Gamma^4}, \qquad \Gamma \ll 1.
\end{equation}
Note that the ``sinh'' in Eq.~\eqref{lin}, for $\omega <\omega_J$,
has now been replaced by a ``sin'' in Eq.~\eqref{linear}, for
$\omega >\omega_J$.

\begin{figure}
\begin{center}
\includegraphics*[width=8cm]{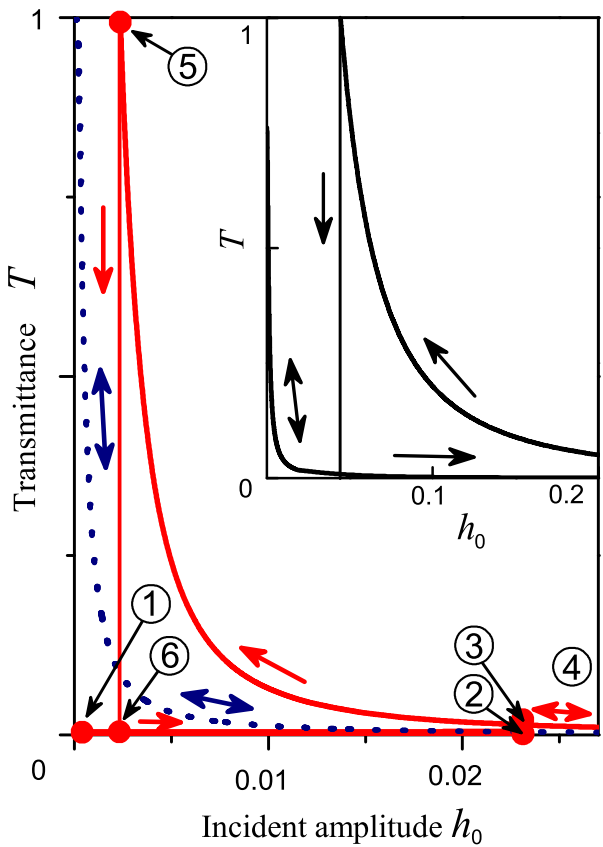}
\caption{(Color online) Dependence of the transmittance $T$ on the
normalized amplitude $h_0$ of the incident wave for different
\emph{positive} values of the frequency detuning: $\Omega-1=
5\cdot 10^{-3}$, or $\delta /\pi = 1.2$ (dotted curve);
$\Omega-1=4.5\cdot 10^{-3}$, or $\delta /\pi = 1.25$ (solid
curve); $\Omega-1=1.65\cdot 10^{-3}$, or $\delta /\pi = 2.1$
(inset). Arrows show the change of the transmittance when changing
$h_0$. The sample thickness is $D=4.3 \cdot 10^{-5}$~cm, and other
parameters are the same as in Fig.~\ref{fig t(h) omega<1}.}
\label{fig t(h) omega>1}
\end{center}
\end{figure}

In the nonlinear case, changing the amplitude $h_0$, one can
control the relation between the wavelength and the thickness of
the slab and, thus, the \emph{transmittance is tunable} by the
amplitude of the incident wave. Figure~\ref{fig t(h) omega>1}
shows the $T(h_0)$ dependences for different positive frequency
detunings.

The analysis based on Eqs.~\eqref{h(a)} (with choice of the sign
``$-$''), \eqref{sine-Gordon}, \eqref{upper}, \eqref{lower ampl},
\eqref{lower deriv}, and~\eqref{R} shows that the dependence
$T(h_0)$ is reversible when the frequency detuning is larger than
some threshold value, which is defined by the asymptotic equation
\begin{equation}\label{thr}
(\Omega_{\rm thr}-1)\, \approx \, \left(\frac{D\sqrt{\epsilon}\sin
\theta }{\sqrt{2}\pi\lambda_{ab}}\right)^2, \quad \Gamma \ll 1.
\end{equation}
An example of such a reversible $T(h_0)$ dependence is presented
by the dotted curve in Fig.~\ref{fig t(h) omega>1}.

The hysteresis in the $T(h_0)$ dependence appears for frequency
detunings smaller than the threshold value:
\begin{equation}\label{hyst}
\Omega < \Omega_{\rm thr}.
\end{equation}
In this case, the transmittance can reach the value one when the
incident wave amplitude $h_0$ is first increased and then
decreased. Namely, the incident wave amplitude $h_0$ is decreased
after it increases and a jump of $T(h_0)$ occurs from the
low-amplitude branch to the high-amplitude one (see the solid
curve and the inset in Fig.~\ref{fig t(h) omega>1}). One can
derive the asymptotic equation for the optimal value $h_0^{\rm
max}$ of $h_0$ when the superconducting slab becomes totally
transparent,
\begin{equation}\label{higla2}
h_0^{\rm max}\simeq\frac{3\sqrt{3}}{4I^2}\Gamma \delta^2
\end{equation}
where
\[
I=\int \limits_0^1
\frac{dx}{\sqrt{1-x^{4/3}}}\,=\,\frac{3}{4}\,\rm B
\Big(\frac{1}{2},\frac{3}{4}\Big)\,\approx \, 1.7972
\]
and ${\rm B}(x,y)$ is the Euler integral of the first kind.

\subsection{Mechanical analogy revisited}

\begin{figure}
\begin{center}
\includegraphics*[width=8cm]{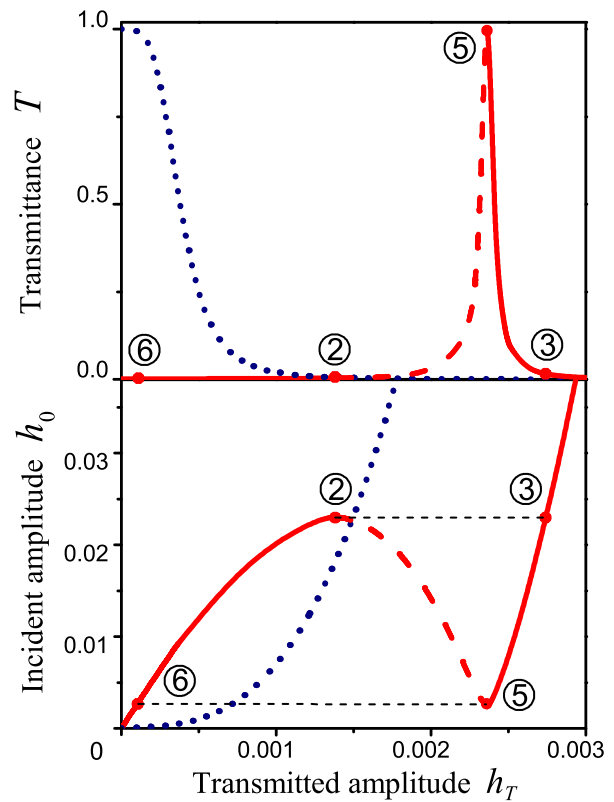}
\caption{(Color online) Solution of the inverse problem: dependences of the amplitude $h_0$ of the
incident wave and transmittance $T=h_T^2/h_0^2$ on the amplitude
$h_T$ of the transmitted wave. The values of the parameters
and the numbers near the indicated points are the same as in the main panel in Fig.~\ref{fig
t(h) omega>1}. The dependences plotted by dotted curves are monotonic, leading
to the single-valued dependence $T(h_0)$ (the dotted curve in
Fig.~\ref{fig t(h) omega>1}). The solid-and-dashed curves show the
nonmonotonic behavior that results in the multivalued
dependence $T(h_0)$  (the solid curve in the main
panel in Fig.~\ref{fig t(h) omega>1}).} \label{fignonmon}
\end{center}
\end{figure}

Returning to the mechanical analogy described in the previous subsection, we note that, in the case of positive-frequency detuning, the dependence of the potential $U_{\rm eff}$ on the radial coordinate $h$ of the particle is single-valued  (see the inset in Fig.~\ref{fig_ueff_<1}). This is because the dependence $a(h)$ in Eq.~\eqref{h(a)} is single-valued in this case. Nevertheless, the dependence $T(h_0)$ can be multivalued even for $\Omega > 1$ (see Fig.~\ref{fig t(h) omega>1}). This feature seems to be paradoxical. Indeed, the particle motion is completely defined for any \emph{initial conditions}.  However, an assignment of the value of $h_0$ in relations \eqref{upper}, \eqref{lower ampl}, and \eqref{lower deriv} \emph{does not mean an imposition of definite initial conditions} for the particle motion. To illustrate this nontrivial feature of the electromagnetic wave transmission through a slab of layered superconductor, let us now consider the inverse problem. We wish to find the amplitude $h_0$ of the incident wave that is necessary to obtain a given value $h_T$ of the transmitted wave. According to Eq.~\eqref{lower ampl}, the value of $h_T$ defines unambiguously the angular momentum $L=h_T^2/\Gamma$ of the particle. On the basis of the motion equation \eqref{sine-Gordon} and the boundary condition Eq.~\eqref{upper}, we find that the dependence $h_0(h_T)$ should be single-valued. Correspondingly, the dependence of the transmittance
\[
T=h_T^2/h_0^2
\]
on the amplitude $h_T$ of the transmitted wave is also single-valued (see Fig.~\ref{fignonmon}).  However, \emph{this dependence is nonmonotonic} if the condition Eq.~\eqref{hyst} is satisfied. As a result, the dependence $T(h_0)$ appears to be multiple-valued.

\section{Conclusion}

In this paper we predict a new nonlinear phenomenon in layered
superconductors. We show that the reflectivity and transmittance
of a superconducting slab are very sensitive to the amplitude of
the incident wave due to the nonlinearity of the Josephson
relation for the \textbf{c}-axis current. As a result, the
reflectivity and transmittance vary over a wide range, from nearly
zero to one (if neglecting the small dissipation), when changing
the amplitude of the incident electromagnetic wave. A remarkable
feature of this phenomenon is the hysteretic behavior of the
$T(h_0)$ dependence. It is important to note that, for frequencies
close to the Josephson plasma frequency, the tunable transmittance
can vary from zero to one even in the case of weak nonlinearity,
when the interlayer phase difference $\varphi$ is small, $\varphi
\ll 1$.

\section{Acknowledgements}

We gratefully acknowledge partial support from the Laboratory of Physical Sciences, National Security Agency, Army Research Office, National Science Foundation grant No. 0726909, DARPA, JSPS-RFBR contract No. 09-02-92114, Grant-in-Aid for Scientific Research (S), MEXT Kakenhi on Quantum Cybernetics, Funding Program for Innovative R\&D on S\&T (FIRST), and Ukrainian State Program on Nanotechnology and Nanomaterials.

\end{document}